\documentstyle[12pt]{article}
\begin{document}
\renewcommand{\thesection}{\Roman{section}}
\baselineskip 20pt
\input feynman.tex
{\hfill PUTP-96-08}
\vskip 3cm
\begin{center}
{\large\bf Color-Octet Charmonium Production in Top Quark Decays}
\end{center}
\vskip 7mm
\centerline{Cong-feng Qiao~~~Kuang-Ta Chao}
\centerline{\small\it Center of Theoretical Physics, CCAST(World Laboratory), 
            Beijing 100080, P.R. China}
\centerline{\small\it Department of Physics, Peking University, 
            Beijing 100871, P.R. China}
\begin{center}
\begin{minipage}{120mm}
\vskip0.6in
\begin{center}{\bf Abstract}\end{center}
  {We calculate the direct production rate of $J/\psi$ in top quark decays. 
  The color-octet $J/\psi$ 
  production via $t\rightarrow W^+ b J/\psi$ is 
  shown to have a large branching ratio of order $1.5\times 10^{-4}$, 
  which is over an order of magnitude higher than that of the color-singlet 
  $J/\psi$ production via $t\rightarrow W^+ b J/\psi~g~g $ or $t\rightarrow    
  W^+ b~\chi_{cJ}~g$ followed by $\chi_{cJ}\rightarrow J/\psi~\gamma$. This  
  result can be used as a powerful tool to test the importance of the 
  color-octet mechanism in heavy quarkonium production.}
\vskip 1cm
PACS number(s):$14.40. Lb$~ $14.65. Ha$
\end{minipage}
\end{center}

\vfill\eject\pagestyle{plain}\setcounter{page}{1}
Since the discovery of charmonium in 1974, there have been a lot of attempts
to interpret the production of 
these new states. Among many scenarios, the color-singlet model\cite{1} 
gains more success than other alternatives\cite{2} like the color-evaporation 
model\cite{1}\cite{3}. Based upon the color-singlet model,  
it is possible to calculate the production rates
from first principles by standard methods\cite{4}. 
Indeed, the  study of heavy quarkonium production may provide a suitable 
ground to precisely test 
quantum chromodynamics (QCD). 

However, during the past 
few years, it is found that the color-singlet model also has
some defects in describing the production of heavy quarkonium. 
Predictions for the S-wave charmonium production failed 
to explain the new data. In the 1992-1993 run,
the {\bf CDF} detector at Fermilab Tevatron\cite{5} gave rates for prompt 
$\psi$ and $\psi^{\prime}$ production at large transverse momentum 
which were orders of mgnitude above the lowest order perturbative calculation
within the color-singlet model\cite{6}. Even after including the 
fragmentation contributions\cite{7}\cite{8}\cite{9}which overwhelm the former
when $P_{T}\ge 6GeV$\cite{9p}, there is still left a big gap between theory and 
experiment. 

Recently, a new effective field theory for bound states of heavy quark and 
antiquark was provided by Bodwin, Braaten, and Lepage\cite{10} in the context
of non-relativistic quantum chromdynamics(NRQCD). In this approach, the 
interaction operators are expanded in powers of $ v $, the velocity
of heavy quark and antiquark in the meson rest frame, and $\alpha_s$.
Contributions of different orders in $ v$ are seperated according to the
``velocity scaling'' rules. 
In this new framework a heavy quarkonium state $H$ is not solely
regarded as simply a quark-antiquark pair but rather a superposition of 
a series of Fock states:  
\begin{eqnarray}
\nonumber
|H(nJ^{PC})\rangle&=&O(1)|Q\bar{Q}(^{2S+1}L_J,\b{1})\rangle\\
\nonumber
                   &+&O(v)|Q\bar{Q}(^{2S+1}(L\pm 1)_{J^{\prime}},\b{8})\rangle\\
\nonumber
                   &+&O(v^2)|Q\bar{Q}(^{2S+1}L_J,\b{8}~or~\b{1})gg\rangle\\
                   &+&\cdots,
\end{eqnarray}
where the angular momentum of the $Q\bar{Q}$ pair in each Fock state is 
labeled by$~^{2S+1}L_J$ with a color configuration of either $\underline 8$
or $\underline 1$. The pure $Q\bar{Q}$ state in color-singlet is only 
the leading term in the above expansion. Up to and including ${\cal O}(v^2)$
in the Fock state expansion in describing $J/\psi(\psi^{\prime})$
production, the color-octet matrix element 
$<{\cal O}^{J/\psi}_8(^3S_1)>[<{\cal O}^{\psi^\prime}_8(^3S_1)>]$ should also
be taken into consideration. Although these color-octet matrix elements are 
suppressed by order of $ v^4$ relative to the corresponding 
color-singlet matrix elements 
$<{\cal O}^{J/\psi}_1(^3S_1)>[<{\cal O}^{\psi^\prime}_1(^3S_1)>]$,
they are enhanced by a factor of $1/{\alpha_s^2}$ relative to the 
color-singlet process in the short-distance perturbative calculatuion. 
Therefore, the supperession in the color-octet matrix elements can be 
compensated. Treating the color-octet matrix elements as free 
paramenters, the description of high-$P_T~J/\psi(\psi^{\prime})$ 
production at the Tevatron can 
indeed be rescued\cite{11}\cite{12}\cite{13}, but clearly more work is needed 
before the new formalism is established as a successful theory of 
quarkonium production. To this end, a list of papers have been published 
\cite{14}\cite{15}\cite{16}, and still in this paper, we  suggest using 
another important process to test the color-octet quarkonium 
production mechanism.
 
The success of the Standard Model (SM)\cite{17}\cite{18}\cite{19} suggests 
that the top quark must 
exit \cite{20}. Recently, from the direct search at the Tevatron, 
the {\bf CDF} and {\bf D0} groups confirmed the existence of a heavy 
top quark\cite{21}\cite{22}, with a mass of $(176 \pm 8 \pm 10) $ GeV or
$(199^{+19}_{-21} \pm 22 )$ GeV. The next experimental studies will 
focus on the determination of its properties.
Among others, the measurement of top quark decays into heavy quark mesons
which are made of charm or bottom quark and antiquark, will be of special
interest. In particular, the charmonium production in top quark decays will
provide very useful information in testing the Standard Model.

In the Standard Model, charmonium(e.g. $J/\psi$) may be produced via the 
flavor changing neutral current transition $t\rightarrow c g$\cite{tcg} 
followed by
the charm quark fragmentation $c\rightarrow J/\psi c$ or the gluon 
fragmentation $g\rightarrow J/\psi$, but the rates of these processes are 
very small within the Standard Model and hence sensitive to the new 
physics beyond the Standard Model\cite{cqy}.
On the other hand, however, since
in the Standard Model the dominant decay mode of top quark is 
$t\rightarrow W^+ b$\cite{24}, the dominant direct charmonium production is
expected to proceed via $t\rightarrow W^+ b g^*$ with the virtual gluon $g^*$
fragmentation into charmonium. There are three subprocesses for $J/\psi$
production via $t\rightarrow W^+ b g^*$: ($i$) color-octet gluon fragmentation 
$g^*\rightarrow J/\psi$; ($ii$) color-siglet gluon fragmentation 
$g^*\rightarrow J/\psi g g $; ($iii$) color-singlet gluon fragmentation
$g^*\rightarrow \chi_{cJ} g$ followed by $\chi_{cJ}\rightarrow J/\psi \gamma$,
where $\chi_{cJ}(J=0,1,2)$ are the $P$-wave states.

The direct charmonium production appears  
at orders $\alpha_s^4$ or over in the color-singlet model\cite{25}, whereas
the color-octet production process  given by $t\rightarrow W^+ b J/\psi$, 
as shown in Fig.1, is at order $\alpha_s^2$. Its amplitude may be written as
\begin{eqnarray}
{\cal A}(t\rightarrow W^+ b J/\psi)
&=& \frac{i g g^2_s V_{tb}}{2\sqrt{2}M} T^{a}\epsilon^{\mu}_{\psi}
\epsilon^{\alpha}_{W} \bar{u}(P^\prime)
\Big[\gamma_\mu \frac{1}{{\not\! P}^\prime
+ \not\!P - m_b}\gamma_\alpha (1-\gamma_5)\\
& +& \gamma_\alpha (1-\gamma_5)
\frac{1}{{\not\!P}^\prime
+ \not\! K -m_t}\gamma_\mu \Big] u(T) {\cal M}_8(J/\psi),
\end{eqnarray}
where $T,~P',~K$, and $P$ are the 4-momenta of the top quark, $b$ quark, 
$W^+$ boson, and $J/\psi$ respectively, and 
${\cal M}_8(J/\psi)\equiv |{\cal A}(Q\bar{Q}[^3S_1^{(8)}]\rightarrow 
J/\psi)|$ is the long distance nonperturbative amplitude of all possible way
of evolving to $J/\psi$ starting from a color-octet 
$Q\bar{Q}[^3S_1^{(8)}]$ pair at short distances.
It may be treated as phenomenological parameter which can be determined 
by fitting the data, e.g. from the $J/\psi$ production rate at 
the Tevatron\cite{13}.
We define
\begin{eqnarray}
\nonumber
&&f_1 = -(M^2 + 2 m_t^2) [(m_t^2 - m_b^2)^2 + (m_t^2 + m_b^2 - 2 m_w^2) m_w^2]~, \\
\nonumber      
&&f_2 = - 2 (m_t^2 - m_b^2)^2 (M^2 + m_b^2 + m_t^2) 
 + 2 m_w^2 [(2 M^2 + 3 m_w^2)(m_b^2 + m_t^2)\\ 
\nonumber  
&&~~~~~~~ - 2 (M^2 + m_w^2)^2 - 4 m_b^2 m_t^2]~,~~~~~~f_3 =  -m_b^2 - m_t^2 - 2 m_w^2 ~,\\ 
\nonumber      
&&f_4 = 2 m_b^4 - 4 m_b^2 m_t^2 + 2 m_t^4 - 4 M^2 m_w^2  + 
      2 m_b^2 m_w^2 + 2 m_t^2 m_w^2 - 4 m_w^4 ~, \\ 
      \nonumber
&&f_5 = -(M^2 + 2 m_b^2) [(m_t^2 - m_b^2)^2 + (m_t^2 + m_b^2 - 2 m_w^2) m_w^2 ]~,\\
\nonumber      
&&f_6 = 2 (m_t^2 - m_b^2)^2 - 2(2 M^2 - m_b^2 - m_t^2 + 2 m_w^2) m_w^2 ~,\\ 
\nonumber      
&&f_7 = - 2 m_b^2 - 2 m_t^2 ~,~~~~f_8 = - m_b^2 - m_t^2 - 2 m_w^2 ~,
\end{eqnarray}
where $M=2 m_c$ is the $J/\psi$ mass. 
Then, the differential decay rate for $t\rightarrow W^+ b J/\psi$ is given by
\begin{eqnarray}
\label{dd}
\frac{d^2\Gamma}{dx_1 dx_2}(t\rightarrow W^+ b J/\psi)&=& 
    \frac{g^2\alpha_s^2 |V_{tb}|^2 |{\cal M}_8(J/\psi)|^2}
    {24 \pi M^2 m_t^3 m_w^2}\{ f_1 x_1^2 + f_2  x_1 x_2 +
    f_3 x_1^3 x_2 + f_4 x_1^2 x_2  \nonumber \\
    &+&  f_5 x_2^2 + f_6 x_1 x_2^2 + f_7 x_1^2 x_2^2 + 
    f_8 x_1 x_2^3\}/(x_1^2 x_2^2)~.
\end{eqnarray}
Here, the variables $x_1=m_b^2 - m_t^2 - m_w^2 + 2 m_t E_w,
~x_2=2 m_t E_{J/\psi} - M^2$. 
The physical limits of $x_1$ and $x_2$ are
\begin{eqnarray}
\nonumber
x_1^{\pm}\! & = &\!\frac{1}{2(m_t^2 - x_2)}
  \{(M^2 - x_2)(m_t^2+m_b^2-m_w^2-x_2) \\
\nonumber
&\pm &\!\lambda^{\frac{1}{2}}[(m_t^2 - x_2),m_t^2,M^2]
\lambda^{\frac{1}{2}}[(m_t^2 - x_2),m_b^2,m_w^2]\} - M^2,\\
&&{x_2^{-}=m_t^2-(m_t-M)^2,~~~x_2^{+}=m_t^2-(m_b+m_w)^2}.
\end{eqnarray}
Here $\lambda(x,y,z)\equiv(x-y-z)^2-4yz$. 
Setting $\alpha_s=0.253,~m_c=1.5GeV,~m_b=4.9GeV,~
m_t=176GeV$\cite{11}, and $|{\cal M}_8(J/\psi)|^2=0.68\times 10^{-3}~GeV^2$
\cite{13}, we get the branching ratio of 
\begin{eqnarray}
B(t\rightarrow W^+ b J/\psi)\approx 1.46\times 10^{-4}. 
\end{eqnarray}

The dominant color-singlet prompt $J/\psi$ production process to be
$t\rightarrow W^+ b g^*$ with $g^*\rightarrow J/\psi gg$, and 
$g^*\rightarrow \chi_{cJ} g$ followed by $\chi_{cJ}\rightarrow J/\psi \gamma$, 
as shown in Fig.2. We can estimate the partial width following the way in
Ref.\cite{26}. The differential decay rate of $t\rightarrow W^+ b g^*$
is similar to  Eq.(\ref{dd}), and can be easily obtained 
or found in Ref.\cite{ad}. With the definition
\begin{eqnarray}
\label{lll}
\Gamma(g^*\rightarrow A X)= \pi \mu^3 P(g^*\rightarrow A X),
\end{eqnarray}
the decay distribution $P(g^*\rightarrow \chi_{cJ} g)$ and 
$P(g^*\rightarrow J/\psi gg)$ for the gluon of 
virtuality $\mu$ can be found in Ref.\cite{4} and Ref.\cite{27}.
\begin{eqnarray}
&&\mu^2 P(g^*\rightarrow \chi_{c0} g)=\frac{r(1-3 r)^2}{1-r} C_p,\\
&&\mu^2 P(g^*\rightarrow \chi_{c1} g)=\frac{6r(1 + r)}{1-r} C_p,\\
&&\mu^2 P(g^*\rightarrow \chi_{c2} g)=\frac{2r(1 + 3 r + 6r^2)}{1-r} C_p,\\
\label{tt1}
&&\mu^2 P(g^*\rightarrow J/\psi gg)=
C_s r \int\limits_{2\sqrt{r}}^{1+r} dx_{J/\psi}
 \int\limits_{x_{-}}^{x_{+}}dx_{1} f(x_{J/\psi},x_1;r),
\end{eqnarray}
where $r\equiv M/{\mu}$, M is the mass of the relevant charmonium states, and
\begin{eqnarray}
\label{uu}
C_p=\frac{8\alpha_s^2}{9\pi}\frac{|R_p^\prime(0)|^2}{M^5},~~~
C_s=\frac{5\alpha_s^3}{27\pi^2}\frac{|R_s(0)|^2}{M^3} .
\end{eqnarray}
The function $f$ in Eq.(\ref{tt1}) is of the form\cite{27}
\begin{eqnarray}
\nonumber
f(x_{J/\psi},x_1;r)&=&\frac{(2+x_2)x_2}{(2-x_{J/\psi})^2(1-x_1-r)^2} +
 \frac{(2+x_1)x_1}{(2-x_{J/\psi})^2(1-x_2-r)^2}\\ 
 \nonumber
 &+&
 \frac{(x_{J/\psi}-r)^2-1}{(1-x_2-r)^2(1-x_1-r)^2}
+ \frac{1}{(2-x_{J/\psi})^2}\Big(\frac{6(1+r-x_{J/\psi})^2}
 {(1-x_2-r)^2(1-x_1-r)^2}\\ 
& + & \frac{2(1-x_{J/\psi})(1-r)}{(1-x_2-r)(1-x_1-r)r}
  +\frac{1}{r}\Big ),
\end{eqnarray}
where 
$x_{i}\equiv 2E_{i}/\mu$ with $i=J/\psi,~g_1,~g_2$ are the energy fractions
carried by the $J/\psi$ and two gluons in the $g^*$ rest frame, and then 
$x_2 = 2 - x_1 - x_{J/\psi}$.
The limits of the $x_1$ integration in Eq.(\ref{tt1}) 
are
\begin{eqnarray}
x_{\pm}= \frac{1}{2}(2 - x_{J/\psi} \pm \sqrt{x_{J/\psi}^2-4 r}) .
\end{eqnarray}
We can evaluate the total decay rate of top quark to various color-singlet
charmonium states, $A$, via
\begin{eqnarray}
\Gamma(t\rightarrow W^+ b g^*;~g^*\rightarrow A X)=\int\limits_{M^2}^{m_t^2/4} 
d\mu^2 \Gamma(t\rightarrow W^+ b g^*(\mu^2))\cdot P(g^*\rightarrow A X).
\end{eqnarray}
In the numerical estimation, we take 
$\alpha_s= 0.253,~m_c = 1.5 GeV,~M= 2 m_c,~
|R_s(0)|^2=0.999~GeV^3$, and $|R_p^\prime (0)|^2=0.125~GeV^5$\cite{28}, and get
\begin{eqnarray}
&&B(t\rightarrow W^+ b \chi_{c0} g)\cdot B(\chi_{c0}\rightarrow J/\psi \gamma)=2.49\times 10^{-9},\\
&&B(t\rightarrow W^+ b \chi_{c1} g)\cdot B(\chi_{c1}\rightarrow J/\psi \gamma)=5.35\times 10^{-6},\\
&&B(t\rightarrow W^+ b \chi_{c2} g)\cdot B(\chi_{c2}\rightarrow J/\psi \gamma)=1.88\times 10^{-6},\\
&&B(t\rightarrow W^+ b J/\psi gg)=1.39\times 10^{-6}.
\end{eqnarray}
The $\chi_{cJ}$ production rates depend on the infrared cutoff. Here we take
the cutoff $\mu^2_{min}=2 M^2$, which is the same as that in 
the fragmentation 
analysis\cite{29}. Adding the branching ratios together, we obtain the total
color-singlet prompt $J/\psi$ production rate to be 
$8.6\times 10^{-6}$, which is about 
a factor of $20$ smaller than that via the color-octet production mechanism.

In conclusion, we have considered the color-octet charmonium 
production in top 
quark decays, and found the branching ratio of this dominant process 
$t\rightarrow W^+ b J/\psi$ to be $1.46\times10^{-4}$, which is over an order 
of magnitude larger than that of 
color-singlet production processes. Such a large difference makes 
the process of charmonium production in top decay another important channel 
to identify color-octet qurkonium signals whenever there are enough top quark 
events at the Fermilab Tevatron, LHC (Large Hadron Collider), or 
NLC (Next Linear Collider) in the future.
\vskip 1cm
\begin{center}
\bf\large\bf{Acknowlegement}
\end{center}

This work was supported in part by the National Natural Science Foundation
of China, the State Education Commission of China and the State Commission
of Science and Technology of China.

\newpage
\centerline{\bf \large Figure Captions}
\vskip 2cm
\noindent Fig.1.  Color-octet Charmonium production process in top quark decays.\\

\noindent Fig.2.  Diagrams for color-singlet $J/\psi$ production. 
(a)via $g^*\rightarrow J/\psi g g$ (b)
via $g^*\rightarrow \chi_{cJ} g \rightarrow J/\psi\gamma g$. For diagram(a)
$x_{i}\equiv 2E_{i}/\mu$ with $i=J/\psi,~g_1,~g_2$ are the energy fractions
carried by the decay products in $g^*$ rest frame normalized to 
$\mu\equiv m(g^*)$.
\end{document}